\documentclass[prl,letterpaper,preprint]{revtex4-1}
\usepackage[utf8x]{inputenc}
\usepackage{amsmath}
\usepackage{graphicx}
\usepackage{dcolumn}
\usepackage{bm}
\usepackage{subfigure}

\newcommand{\ket}[1]{\left| #1 \right\rangle}
\newcommand{\bra}[1]{\left\langle #1 \right|}

\newcommand{\dpar}[2]{\frac{\partial #1}{\partial #2}}

\newcommand{\dx}[1]{{\rm d} #1 \,}

\begin{document}
\begin{abstract}
We calculate the bulk photovoltaic response of the ferroelectrics BaTiO$_3$ and PbTiO$_3$ from first principles by applying ``shift current" theory to the electronic structure from density functional theory.  The first principles results for BaTiO$_3$ reproduce eperimental photocurrent direction and magnitude as a function of light frequency, as well as the dependence of current on light polarization, demonstrating that shift current is the dominant mechanism of the bulk photovoltaic effect in BaTiO$_3$.  Additionally, we analyze the relationship between response and material properties in detail.  The photocurrent does not depend simply or strongly on the magnitude of material polarization, as has been previously assumed; instead,  electronic states with delocalized, covalent bonding that is highly asymmetric along the current direction are required for strong shift current enhancements.  The complexity of the response dependence on both external and material parameters suggests applications not only in solar energy conversion, but to photocatalysis and sensor and switch type devices as well. 
\end{abstract}
\title{\bf First principles calculation of the shift current photovoltaic effect in ferroelectrics}
\author{Steve M. Young and Andrew M. Rappe}
\affiliation{The Makineni Theoretical Laboratories, Department of Chemistry, University of Pennsylvania, Philadelphia, PA 19104-6323, USA}

\maketitle

\section{Introduction}
The bulk photovoltaic effect - or photogalvanic effect - refers to the generation of intrinsic photocurrents that can occur in single-phase materials lacking inversion symmetry~\cite{Chynoweth56p705,Chen69p3389,Glass74p233,Fridkin01p654}.  Ferroelectrics (materials that possess intrinsic, switchable polarization) exhibit this effect strongly, producing current in response to unpolarized, direct illumination. Traditionally, photovoltaic materials are heterogeneous, doped structures, relying on the electric field at a $p-n$ junction to separate photoexcited electrons and holes.  By contrast, the bulk photovoltaic effect can be observed even in pure homogeneous samples, as with BaTiO$_3$~\cite{Koch76p305}. Recently, the effect has been demonstrated in the multiferroic BiFeO$_3$, with reported efficiencies as high as 10\%~\cite{Choi09p63,Yang10p143,Seidel11p126805}. Though ferroelectric photovoltaics are currently receiving a great deal of interest, the origins of their photovoltaic properties are considered unresolved. Attention has been focused on interface effects, crystal orientation, and the influence of grain boundaries and defects, while any bulk photovoltaic contributions have been largely ignored~\cite{Qin09p022912,Qin08p122904,Pintilie07p064109,Basu08p091905,Ichiki05p222903,Yue10p015104,Yuan09p252904,Yang09p062909,Pintilie10p114111,Cao10p102104}. Its mechanism has been proposed to be a combination of nonlinear optical processes, especially the phenomenon termed the ``shift current''~\cite{Dalba95p988,vonBaltz81p5590,Tonooka94p224,Sturman1992}, but this has not been firmly established, and the detailed dependence on material properties, especially in ferroelectrics, is not known. This has also hindered progress towards understanding other photovoltaic effects, as the bulk contributation could not be separated out.  While shift current calculations have been performed for some non-ferroelectrics, no experimental comparisons were performed~\cite{Sipe00p5337,Nastos06p035201,Nastos10p235204}.   Here we present the first direct comparison of current computed from first principles with experimentally measured short-circuit photocurrent.  Using the shift current theory, we successfully predict short circuit photocurrent direction, magnitude, and spectral features, demonstrating that shift current dominates the bulk photovoltaic response.  Additionally, we explore the relationship between material polarization and shift current response, making progress towards identifying the electronic structure properties that influence current strength.  

\section{Main}
We emphasize that nonlinear optical processes can give rise to a truly bulk effect~\cite{vonBaltz81p5590,Atanasov96p1703,Tonooka94p224}. The results demonstrate that the most important of these is the
shift current, which arises from the second-order interaction with monochromatic light.  The electrons are excited to coherent superpositions, which allows for net current flow due the asymmetry of the potential. 
Bulk polarization is not required; only inversion symmetry must be broken. Shift currents have been investigated experimentally~\cite{Bieler06p083710,Bieler07p161304,Loata08p1261,Priyadarshi09p151110,Racu09p2784,Rice94p1324,Dalba95p988,Kohli11p470,Ji10p1763}, analytically~\cite{Kraut79p1548,vonBaltz81p5590,Sipe00p5337,Kral00p4851,Kral00p1512}, and computationally, though only for a few nonpolar materials~\cite{Sipe00p5337,Nastos06p035201,Nastos10p235204}. 
Perturbative analysis treating the electromagnetic field classically yields the shift current expression~\cite{vonBaltz81p5590,Sipe00p5337}
\begin{flalign}
  J_q&=\sigma_q^{rs}E_rE_s \nonumber\\
   \sigma_q^{rs}(\omega)  = &\pi e\left(\frac{e}{m\hbar\omega}\right)^2 \sum_{n',n''}\int \dx{\mathbf{k}}\left(f[n'' \mathbf{k}] -f[n' \mathbf{k}] \right)  \nonumber\\ &\times\bra{n' \mathbf{k}} \hat{P}_r  \ket{n'' \mathbf{k}} \bra{n'' \mathbf{k}} \hat{P}_s\ket{n' \mathbf{k}}\nonumber\\&\times\left( -\dpar{\phi_{n' n''}(\mathbf{k}, \mathbf{k})}{k_q} - \left[\chi_{n''q}(\mathbf{k})-\chi_{n'q}(\mathbf{k})\right]\right)\nonumber\\&\times\delta \left(\omega_{n''}(\mathbf{k})-\omega_{n'}(\mathbf{k}) \pm \omega\right) \label{scfermi}
\end{flalign}
where $n',n''$ are band indices, $\mathbf{k}$ is the wavevector in the Brillouin zone, $\omega_n(\mathbf{k})$ is the energy of the $n$th band, so that $\sigma_q^{rs}$ gives the current density response $\mathbf{J}$ to electromagnetic field $\mathbf{E}$, $\chi_{nq}(\mathbf{k})$ denotes the Berry connections for band $n$ at $\mathbf{k}$, and $\phi_{n'n''}$ is the phase of the transition dipole between the bands $n'$ and $n''$. It is worth noting that while the Berry connections introduce a gauge dependence, it is exactly canceled by the gauge dependence of $\dpar{\phi_{n' n''}(\mathbf{k}, \mathbf{k})}{k_q}$, so that the overall expression is gauge invariant.

We may view this expression as the product of two terms with physical meaning
\begin{flalign*}
\sigma_q^{rs}(\omega)=e\sum_{n',n''}\int \dx{\mathbf{k}}I^{rs}(n',n'',\mathbf{k}; \omega)R_q(n',n'',\mathbf{k})
\end{flalign*}
where 
\begin{flalign}
I^{rs}(n',n'',\mathbf{k}; \omega)=&\pi \left(\frac{e}{m\hbar\omega}\right)^2\left(f[n'' \mathbf{k}] -f[n' \mathbf{k}] \right)\nonumber\\&\times\bra{n' \mathbf{k}} \hat{P}_r  \ket{n'' \mathbf{k}} \bra{n'' \mathbf{k}} \hat{P}_s\ket{n' \mathbf{k}}\nonumber\\ &  \times  \delta \left(\omega_{n''}(\mathbf{k})-\omega_{n'}(\mathbf{k}) \pm \omega\right)
\end{flalign}
is the transition intensity, which is proportional to the imaginary part of the permittivity and describes the strength of the response for this transition, and 
\begin{flalign}
 R_q(n',n'',\mathbf{k})=-\dpar{\phi_{n' n''}(\mathbf{k}, \mathbf{k})}{k_q} - \left[\chi_{n''q}(\mathbf{k})-\chi_{n'q}(\mathbf{k})\right]
\end{flalign}
is the shift vector, which gives the average distance traveled by the coherent carriers during their lifetimes.
As an analytical tool, we compute and plot the quantity 
\begin{flalign}
\bar{R}_q(\omega)=&\sum_{n',n''}\int \dx{\mathbf{k}}R_q(n',n'',\mathbf{k})\delta \left(\omega_{n''}(\mathbf{k})-\omega_{n'}(\mathbf{k}) \pm \omega\right).\label{agsv}
\end{flalign}
We note that $\bar{R}$ has units of length over frequency and is not physical, nor is it weighted by intensity; as such the $\bar{R}$ only provides qualitative information about the aggregate shift vector.  
For additional information, see the supplemental materials.

\section{Methods}
Wavefunctions were generated using the Quantum Espresso and Abinit plane-wave Density Functional Theory(DFT) codes with the generalized gradient approximation exchange correlation functional.  Norm-conserving, designed non-local pseudopotentials~\cite{Rappe90p1227, Ramer99p12471} were produced using the Opium package.  Self-consistent calculations were performed on $8\times8\times8$ k-point grids with energy cutoffs of 50~Ry; the resulting 
charge densities were used as input for non-self-consistent calculations on finer k-point grids as necessary. 

\begin{figure}
{
\subfigure[]{     \includegraphics [width=3in]{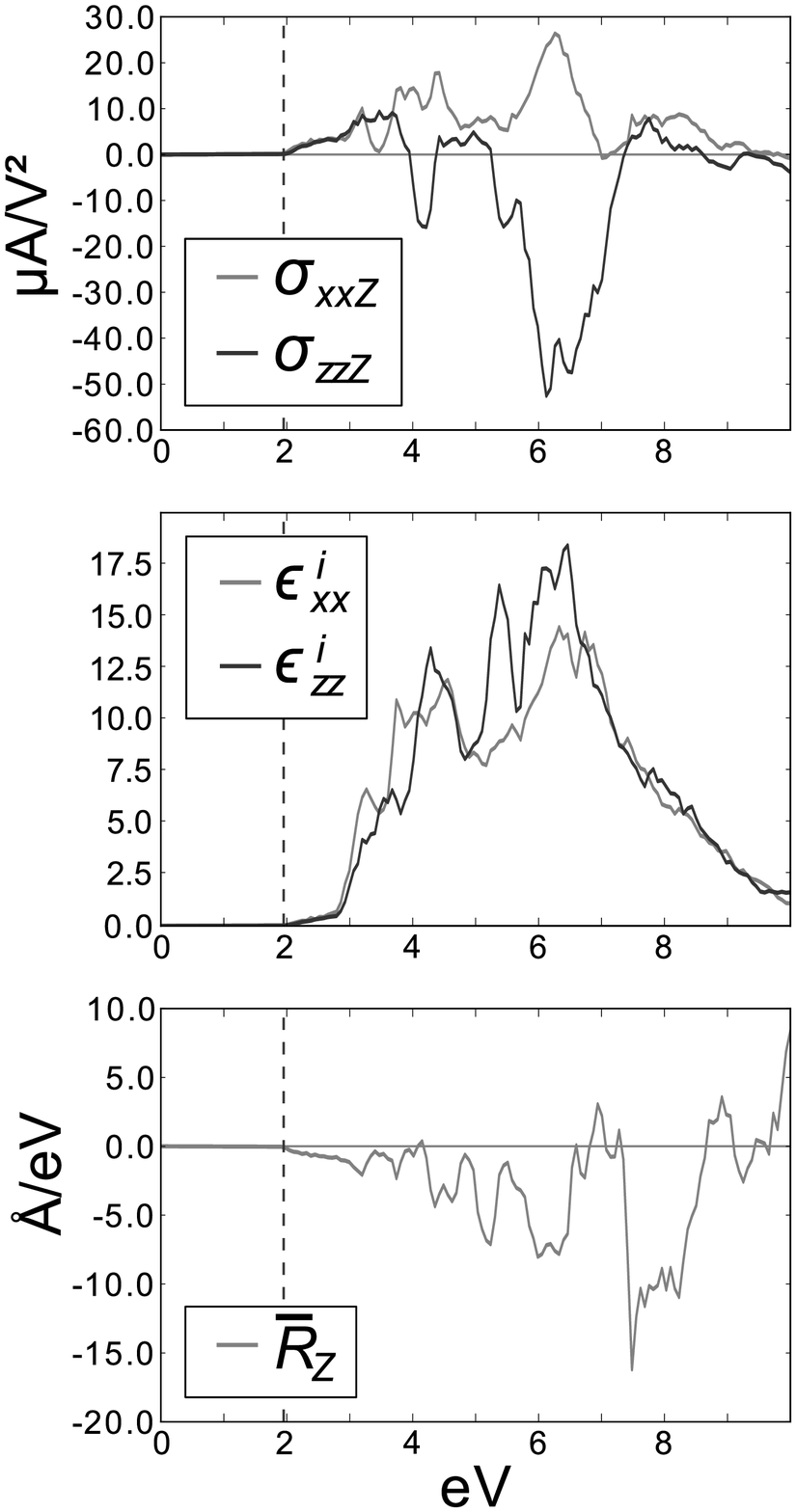}\label{fig:scpt}}
\subfigure[]{     \includegraphics [width=3in]{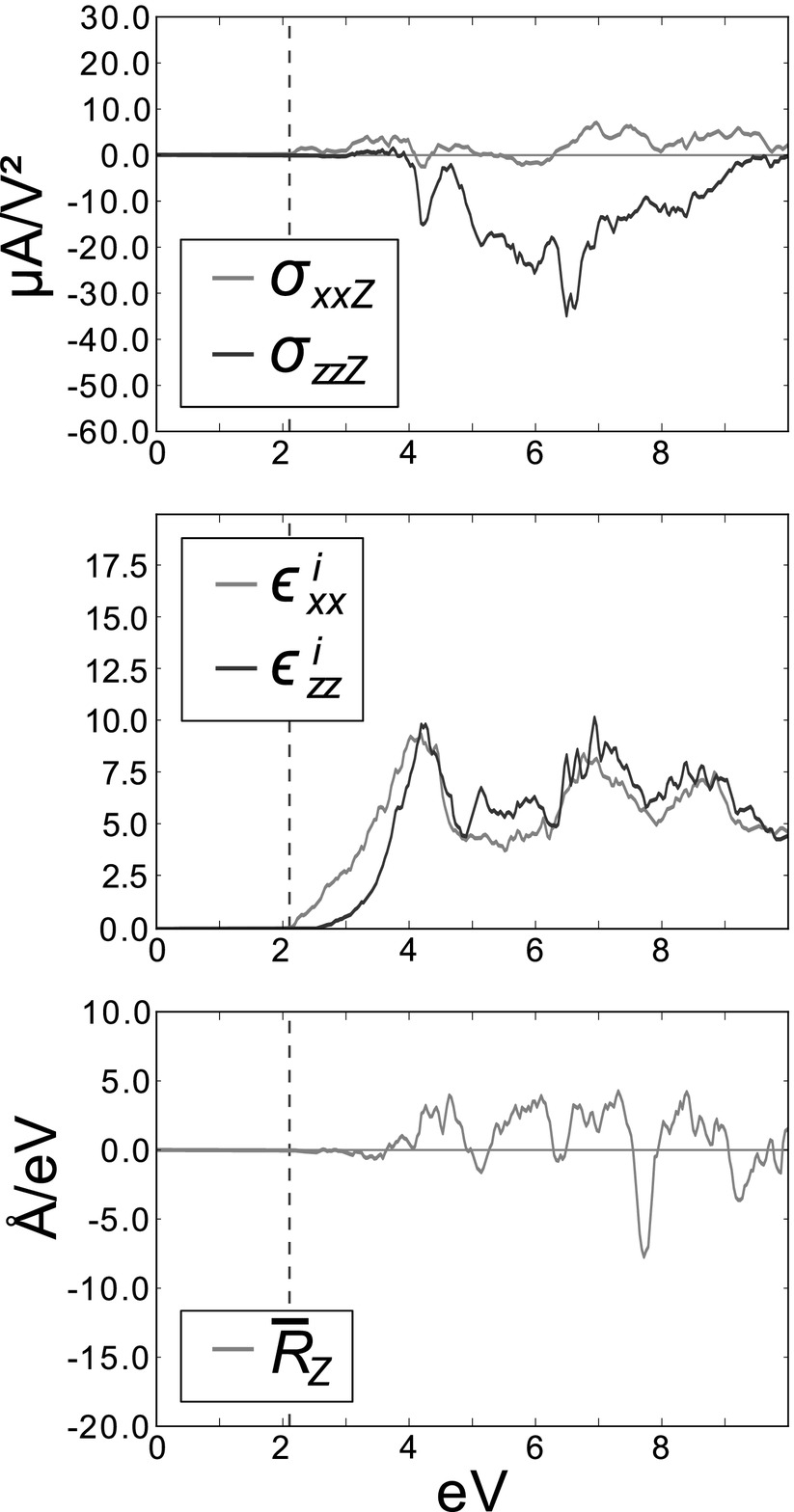}\label{fig:scbt}}

}

\caption{ {The overall current susceptibility~$\sigma$, along with the imaginary component of the permittivity, $\epsilon^i$, and  shift vector integrated over the Brillouin zone~$\bar{R}$, are shown for~\subref{fig:scpt} PbTiO$_3$,~\subref{fig:scbt} BaTiO$_3$. DFT-computed direct band gaps are marked with vertical lines.}}\label{scbtpt}
\end{figure}
\subsection{PbTiO$_3$, BaTiO$_3$}
Lead titanate and barium titanate derive from the cubic perovskite structure and become tetragonal in the ferroelectric phase at room temperature with five-atom unit cells.  Both exhibit strong, robust polarization; combined with their simplicity this makes them ideal candidates for investigating the structural influence on the shift current response.  

The calculations for PbTiO$_3$(PTO) and BaTiO$_3$(BTO) were performed using experimental room temperature geometries~\cite{Glazer78p1065,Buttner92p764}. Shown in Fig.~\ref{scbtpt} are the shift current tensor elements, along with the imaginary component of the permittivity and $\bar{R}_Z$.   Only current response in the direction of material polarization~($Z$) is shown.  The two materials show broadly similar behavior,
with the peak of response several eV above the band gap and well outside the visible spectrum, while the shift current at energies near the band gap is small. The shift current for both materials is stronger in response to incident light polarization parallel to the direction of ferroelectric distortion than when normal to it. 

The shift current depends weakly on the aggregate transition intensities and shift vectors.  In PTO, at 5~eV there is a peak in both the current and intensity, yet the shift vector is relatively small. In fact, despite negative aggregate shift vector at many frequencies, the majority of the current response of PTO to $xx$ polarized light is nonetheless positive.  In BTO, the aggregate shift vector direction is largely positive with a negative shift current response under $zz$ polarized illumination.
This indicates that contributions to response can vary significantly across the Brillouin zone, and suggests that strong correlations between large shift vector and high intensity response are possible but not guaranteed. The product of the aggregate transition intensity and shift vector does not determine even the direction of photocurrent; to find the shift current, it is vital to multiply the transition intensity by its associated shift vector, and then sum over bands and k-points.

\subsection{Experimental Comparison}
\begin{figure}
{
\includegraphics [width=6in]{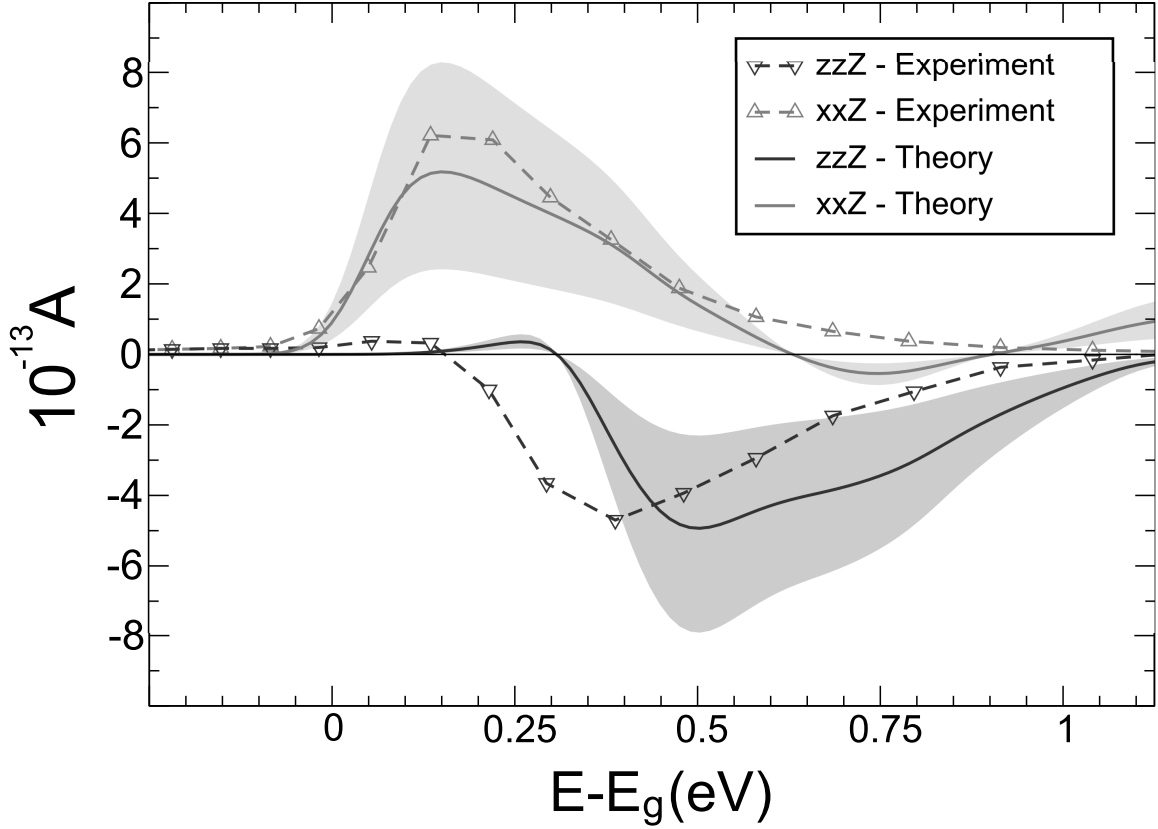}
}

\caption{For BaTiO$_3$, the experimental current~\cite{Koch76p305} and computed current (this work), for transverse~($xxZ$) and longitudinal~($zzZ$) electric field orientation, as a function of energy above their respective bandgaps. The solid lines are calculated results for a choice of experimental parameters of 0.5~mW/cm$^2$ illumination intensity and 0.15~cm sample width.  The shaded regions are bounded by the results using experimental parameters in the given range that provide the lowest and highest response.}\label{expcomp}
\end{figure}

For bulk, single-crystal BaTiO$_3$, experimental spectra are available for energies near the band gap~\cite{Brody75p193,Koch76p305}.  The total current in a bulk crystal for light incident normal to the current direction can by computed from $\sigma_q^{rs}(\omega)$ by
\begin{flalign}
J_q(\omega)=\frac{\sigma^{rr}_q(\omega)}{\alpha^{rr(\omega)}}E^2_rw\\
J_q(\omega)=G^{rr}_q Iw
\end{flalign}
where $\alpha$ is the absorption coefficient, $E$ is the electric field strength, which can be determined from the light intensity $I$, $w$ is the width of the crystal surface exposed to illumination, and $G^{rr}$ is the Glass coefficient~\cite{Glass74p233}.  This expression applies to samples of sufficient thickness to absorb all incident light. We obtained the light intensity and crystal dimensions from~\cite{Koch76p305,Koch75p847}, which were $\approx0.35-0.6$~mW/cm$^2$ and 0.1-0.2~cm, respectively. In Fig.~\ref{expcomp}, the experimental current response from~\cite{Koch76p305} is compared to the response computed using shift current theory.  Despite the uncertainty in experimental parameters, the agreement is striking, in both magnitude and spectrum profile, for both tensor elements. This includes the difference of sign between the majority of the transverse and longitudinal response, which is unusual~\cite{Sturman1992}, as well as the small positive region of the longitudinal response near the band edge.  
For PbTiO$_3$, experimental results suitable for quantitative comparison could not be located.  However, we note that our calculation for PbTiO$_3$ correctly predicts that the current direction is toward the positive material polarization for light frequencies near the band gap~\cite{Ruzhnikov79p49,Daranciang12p087601}, as well as that it's relatively insensitive to light polarization, in contrast to BaTiO$_3$.  

We emphasize that these calculations not only reproduce the magnitude of response, but its idiosyncratic features as well.  Because this theory reproduces all the salient features found in the experiments, this comparison provides strong evidence that shift current is the correct description of the the bulk photovoltaic effect.

\section{Polarization Dependence}
Presently unknown is the relationship of the bulk photovoltaic effect to the material polarization.  Identification of the bulk photovoltaic effect with shift current makes it clear that there is no direct, mechanistic dependence of response on material polarization, as is the case for many mechanisms to which photovoltaic effects in ferroelectrics have been attributed. However, shift current requires broken inversion symmetry, which here derives from the lattice distortion that produces ferroelectric polarization, suggesting that the response may appear to depend on polarization in some fashion.  However, Eq.~\eqref{scfermi} does not reveal a straightforward relationship between the magnitude of symmetry breaking, and the resulting shift current response.  The presented data suggest that stronger polarization does not necessarily imply greater response; photocurrent densities in BaTiO$_3$ and PbTiO$_3$ are of similar magnitude, despite PbTiO$_3$ possessing more than double the material polarization of BaTiO$_3$.
\begin{figure}
{

\includegraphics [width=3in]{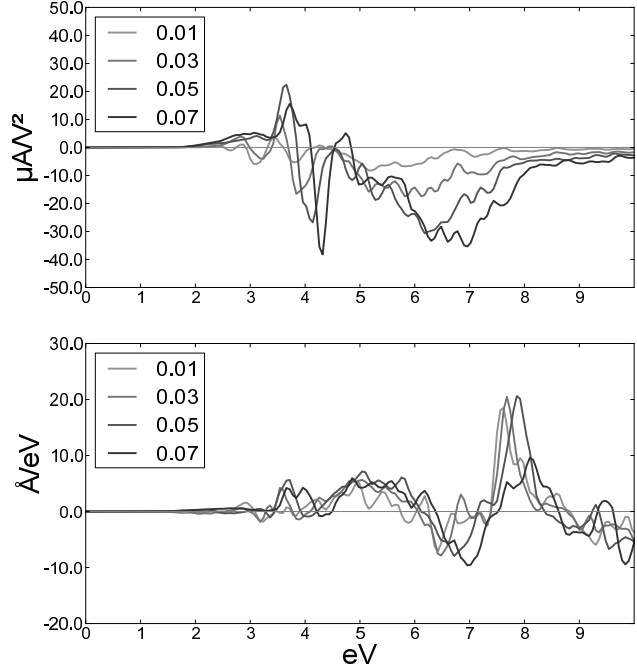}

}
\caption{ \label{scpol}{The overall current susceptibility and aggregated shift vector $\bar{R}$ are shown for PbTiO$_3$ with varying polarization.}}
\end{figure}

To further investigate the connection of photovoltaic effect to material polarization, we studied a systematic family of structures based on PbTiO$_3$.  Starting with the cubic perovskite in the paraelectric structure, we rigidly displaced oxygen ions along a single Cartesian axis by amplitudes ranging from 0.01 to 0.09 lattice vectors, without otherwise altering the geometry.  The spectra of shift current and aggregate shift vector are shown in Fig.~\ref{scpol} for several displacements. The results indicate a complex relationship between shift current and material polarization.  
As Fig.~\ref{scpol} shows, with soft mode amplitude 0.01, the shift current at 3.2
eV above band gap is negative; with amplitude 0.07, the shift current
reverses direction, resulting in a change of -200\%.  With amplitude 0.01,
there is a negative peak at 3.8 eV; with amplitude 0.07, the peak
shifts to 4.2 eV and is four times the size, for an increase of over 300\%. However, in the intervening frequency range, the response is relatively small at all displacements.

Next, we turn our attention to the integrated shift vector~$\bar{R}_Z(\omega)$.  The changes in shift vector are of special interest, since the symmetry constrains the overall shift current expression via the shift vector.  The integrated shift vector spectrum echoes the overall current response, but contains some distinct features.  The increase in current from 4-5~eV does not appear to result from increased shift vector length, but from stronger coincidence of high transition intensity and large shift vectors.  In fact, the overall shift vector changes little with displacement.  However, from 7.5-8.5~eV, the integrated shift vector changes dramatically, suggesting that at some points in the Brillouin zone the oxygen displacement substantially alters the shift vector. Changes to the overall response are thus a combination of changes both to shift vector and associated intensity.

\begin{figure}
{
\subfigure[]{
\includegraphics [width=6in]{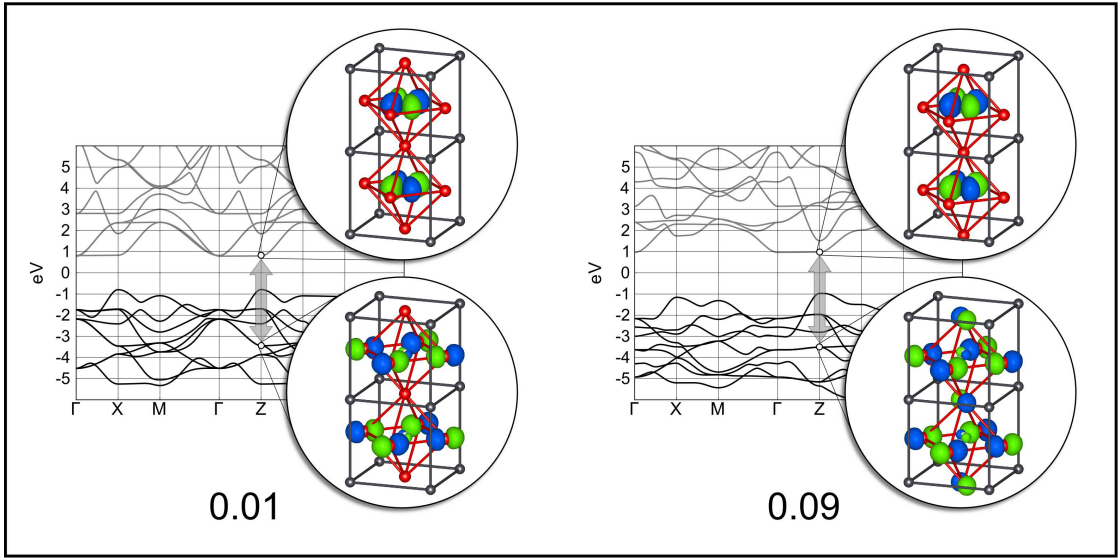}
\label{fig:wflow}
}
\subfigure[]{
\includegraphics [width=6in]{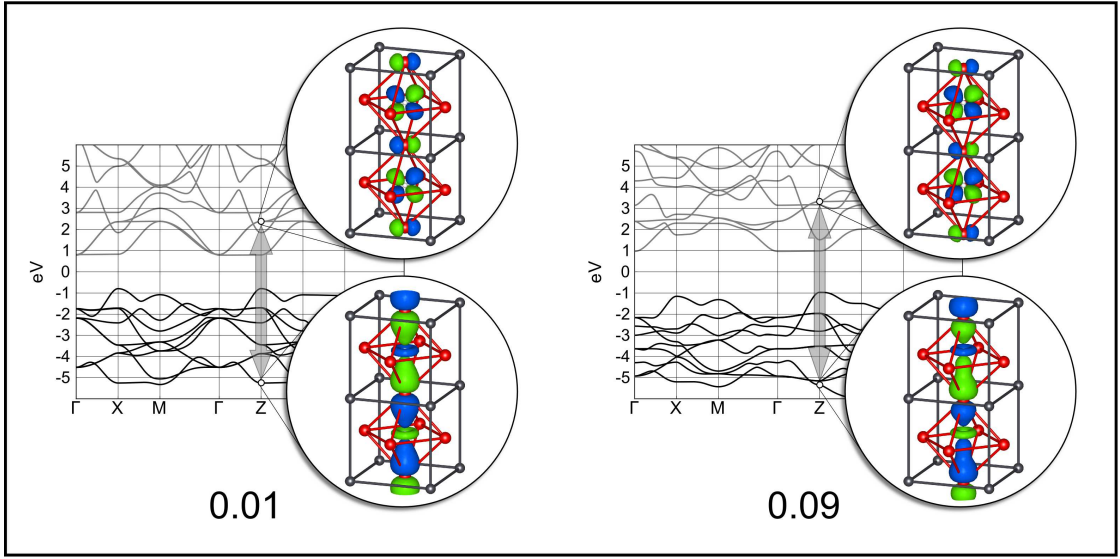}
\label{fig:wfhigh}
}}
\caption{\subref{fig:wflow} The non-bonding Bloch states of PbTiO$_3$ are involved in a transition that is insensitive to material polarization, with a shift vector length change from 0.6~\AA\ to 1.0~\AA\ as O sublattice displacement increases from 0.01 to 0.09, and \subref{fig:wfhigh} a transition from bonding to antibonding gives a shift vector that is highly sensitive to material polarization, with shift vector length change from 32.4~\AA\ to -22.7~\AA\, for increasing O sublattice displacement.}\label{wf}
\end{figure}

To understand these results, the electronic bands participating in transitions in these  frequency ranges were examined directly.  For the 4-5~eV range,  examples of the transitions and associated Bloch states that dominate are shown in Fig.~\ref{wf}\subref{fig:wflow} at 0.01 and 0.09 lattice vector displacements.  For this transition, the shift vector is 0.6~\AA\ at displacement of 0.01, and 1.0~\AA\ at 0.09.  The valence state is largely composed of oxygen $p$-orbitals, while the conduction state is essentially a titanium $d_{xy}$ state.  The states, like the shift vectors, are largely unchanged by the oxygen displacement.

However, the transitions in the higher energy range are notably different.  Shown in Fig.~\ref{wf}\subref{fig:wfhigh} are examples of the dominant transitions in the 7.5-8.5~eV range.  The shift vector is large and positive (32.3~\AA) for 0.01 lattice vector displacement, and large but negative (-22.7~\AA)  at 0.09 displacement.  The participating valence state can be characterized as bonding between the Ti and O atoms collinear with polarization, while the conduction state features Ti-O anti-bonding.  These results point not to a simple dependence on material polarization, but to a dependence of shift current on the extent of localization of the initial and final states, which in turn depends on atomic displacement.  Transitions between states that do not experience bonding interactions in the direction of ferroelectric polarization manifest short shift vectors and insensitivity to oxygen displacement.  

\section{Conclusion}
The shift current response was calculated for ferroelectrics barium titanate and lead titanate.  In the case of barium titanate, the shift current closely matches experiment, successfully predicting magnitude, sign, and spectral profile, including the notable dependence of current direction on polarization.   For lead titanate, where reliable quantitative data are unavailable, the direction and its lack of dependence on polarization were reproduced. This strongly suggests that shift current is the dominant mechanism of the bulk photovoltaic effect in these materials.  

For the materials analyzed, the strongest responses are at frequencies well into the UV spectrum and outside the spectral range probed in most experiments. Consequently, the potential for large shift current response may not yet be fully realized, a conclusion supported by the very large bulk photovoltaic effect observed in response to X-rays~\cite{Dalba95p988}. Furthermore, the strength and direction of the photocurrent subtly depend on the electronic structure of the material, including covalent bonding interactions. This suggests that ferroelectric compounds can vary widely in response profile, and could potentially perform much better than previous results have indicated, encouraging efforts to design materials with large shift current response in the visible spectrum.

\section{Acknowledgments}
The authors thank Gene Mele for valuable discussions. SMY was supported by the Department of Energy Office of Basic Energy
Sciences, under Grant No. DE-FG02-07ER46431.
AMR acknowledges the support of the Office of Naval Research, under
Grant No. N-00014-11-1-0578. Both authors acknowledge computational support from
the HPCMO.

\end{document}


\section{Supplementary Notes}
Here we sketch the derivation of~\eqref{scfermi}, as well as the discretized version suitable for computation
\subsection{Continuous Expression}
In the present work, the optical response is treated semiclassically.   The quantum Liouville equation for the dipole Hamiltonian is expanded to second order.  The perturbation and second-order density matrix term in the interaction picture are
\begin{flalign*}
  \hat{V}(t)  &= \frac{e}{m}\hat{\mathbf{P}}\cdot \hat{\mathbf{A}}_0 \cos \omega t\\
  \hat{\rho}(t)^{I(2)} & =  -\frac{1}{\hbar^{2}}\int^{t}_{0} \int^{t^{\prime}}_{0}\dx{t^{\prime \prime}}\dx{t^{\prime}}\left[\hat{V}(t^{\prime})^{I},\left[\hat{V}(t^{\prime \prime})^{I},\hat{\rho}(0)\right]\right]
\end{flalign*}
The shift current will arise from the $(0; \omega, -\omega)$ terms, corresponding to successive interactions with the positive and negative frequency components of the incident electromagnetic field, creating a zero frequency, or non-oscillating, coherent state.  The current density is  
\begin{flalign}
\mathbf{J} & = \frac{e}{m}\sum_{n',n''}\int \dx{\mathbf{k}} \bra{n' \mathbf{k}'} \hat{\rho}^{I(2)} \ket{n''' \mathbf{k}'''}\bra{n''' \mathbf{k}'''} \hat{\mathbf{P}}^I \ket{n' \mathbf{k}'} \nonumber\\
\mathbf{J} & =  -\pi \left(\frac{e}{m\hbar}\right)^2\frac{e}{m}\sum_{n'\ne n''',n'',u}\int \dx{\mathbf{k}} \Im\left[\bra{n' \mathbf{k}} \hat{\mathbf{P}}\cdot \hat{A}  \ket{n'' \mathbf{k}} \bra{n'' \mathbf{k}} \hat{\mathbf{P}}\cdot\hat{A}\ket{n''' \mathbf{k}} \bra{n''' \mathbf{k}} \hat{\mathbf{P}} \ket{n' \mathbf{k}} \right]\nonumber \\& \qquad\qquad\qquad \times\left[ \left(f[n''' \mathbf{k}]-  f[n'' \mathbf{k}] \right) \delta \left(\omega_{n'''}(\mathbf{k}) - \omega_{n''}(\mathbf{k}) \pm \omega\right)\right. \nonumber\\ &\qquad \qquad\qquad \qquad \qquad \left. -  \left(f[n'' \mathbf{k}] -f[n' \mathbf{k}]  \right)\delta \left(\omega_{n''}(\mathbf{k})-\omega_{n'}(\mathbf{k}) \pm \omega\right)\right] \left(  \frac{1}{\omega_{n'''}(\mathbf{k})-\omega_{n'}(\mathbf{k})} \right)\label{rawcurrent}                                                  
\end{flalign}
The terms where $n'=n'''$ cancel at $\mathbf{k}$ and $-\mathbf{k}$ due to time reversal symmetry, and are excluded.  Thus, only coherent excitations can carry current. 

Using the identity
\begin{flalign}
   \bra{n' \mathbf{k}'} \hat{\mathbf{P}} \ket{n''' \mathbf{k}'''}=-im\left(\omega_{n'''}(\mathbf{k}) -\omega_{n'}(\mathbf{k})\right)\bra{n' \mathbf{k}'} \hat{\mathbf{X}} \ket{n''' \mathbf{k}'''}, n'''\ne n', \mathbf{k}'''\ne \mathbf{k}' \label{pxident}
\end{flalign}
and the position operator for periodic systems first introduced in \cite{Blount62p305},
\begin{flalign}
  \bra{ n'\mathbf{k}'}\hat{\mathbf{X}}\ket{ n\mathbf{k}}&=-i \delta_{nn'}\nabla_{\mathbf{k}}\delta(\mathbf{k}-\mathbf{k}')+\delta(\mathbf{k}'-\mathbf{k}) \int_{\rm WSC} \dx{x} \bra{n'\mathbf{k}'} i\nabla_{\mathbf{k}}\ket{n\mathbf{k}} \label{xop}
\end{flalign}

we can obtain after some manipulation the expression 
\begin{flalign}
  J_q&=\sigma_q^{rs}E_rE_s \nonumber\\
   \sigma_q^{rs}(\omega) & = \pi e\left(\frac{e}{m\hbar\omega}\right)^2 \sum_{n',n''}\int \dx{\mathbf{k}}\left(f[n'' \mathbf{k}] -f[n' \mathbf{k}] \right) \delta \left(\omega_{n''}(\mathbf{k})-\omega_{n'}(\mathbf{k}) \pm \omega\right) \nonumber\\ &\qquad \qquad \qquad\times\bra{n' \mathbf{k}} \hat{P}_r  \ket{n'' \mathbf{k}} \bra{n'' \mathbf{k}} \hat{P}_s\ket{n' \mathbf{k}}\left( -\dpar{\phi_{n' n''}(\mathbf{k}, \mathbf{k})}{k_q} - \left[\chi_{n''q}(\mathbf{k})-\chi_{n'q}(\mathbf{k})\right]\right) \label{scfermi}
\end{flalign}

\subsection{Discrete Expression}
Equation~\eqref{scfermi} cannot be calculated directly on a finite k-point grid.  An appropriate expression must be constructed that retains gauge invariance.  Furthermore, it must accommodate the ambiguity in band identity at and near degeneracies.
Adopting a similar strategy to~\cite{Kingsmith93p1651}, we can reformulate the shift vector using
\begin{flalign*}
  \chi_q &= i\bra{n''\mathbf{k}}\dpar{\,}{k_q}\ket{n''\mathbf{k}} \\
         &=i\iprod{n''\mathbf{k}}{n''\mathbf{k}'}\dpar{\,}{k_q'}\bigg|_{k_q'=k_q}\ln \iprod{n''\mathbf{k}}{n''\mathbf{k}'}\\
         &=i \lim_{\Delta k_q \rightarrow 0}\frac{1}{\Delta k_q} \ln \iprod{n''\mathbf{k}}{n''(\mathbf{k}+\Delta k_q\hat{\mathbf{q}})} -i\ln \iprod{n''\mathbf{k}}{n''\mathbf{k}} \\
  \chi_q &= i \lim_{\Delta k_q \rightarrow 0}\frac{1}{\Delta k_q} \ln \iprod{n''\mathbf{k}}{n''(\mathbf{k}+\Delta k_q\hat{\mathbf{q}})}
\end{flalign*}
and
\begin{flalign*}
 \bra{n'\mathbf{k}}P\left(\dpar{\,}{k_q}\ket{n''\mathbf{k}}\right)&= \bra{n'\mathbf{k}}P\ket{n''\mathbf{k}'}\dpar{\,}{k_q'}\bigg|_{k_q'=k_q} \ln \bra{n'\mathbf{k}}P\ket{n''\mathbf{k}')} \\
    & = \bra{n'\mathbf{k}}P\ket{n''\mathbf{k}} \lim_{\Delta k \rightarrow 0}\frac{1}{\Delta k} \left( \ln \bra{n'\mathbf{k}}P\ket{n''(\mathbf{k} \Delta k_q+\hat{\mathbf{q}})} - \ln \bra{n'\mathbf{k}}P\ket{n''\mathbf{k}}\right) \\
  \bra{n'\mathbf{k}}P\left(\dpar{\,}{k_q}\ket{n''\mathbf{k}}\right)&=\bra{n'\mathbf{k}}P\ket{n''\mathbf{k}} \lim_{\Delta k_q \rightarrow 0}\frac{1}{\Delta k_q} \left( \ln \frac{\bra{n'\mathbf{k}}P\ket{n''(\mathbf{k}+\Delta k_q\hat{\mathbf{q}})}}{ \bra{n'\mathbf{k}}P\ket{n''\mathbf{k}}}\right)
\end{flalign*}
Then we have
\begin{flalign*}
R(n',n'',\mathbf{k}) & =  \lim_{\Delta k_q \rightarrow 0} \frac{1}{\Delta k_q}    i\left(\ln \frac{\bra{n''\mathbf{k}}P\ket{n'\mathbf{k}}\iprod{n'\mathbf{k}}{n'(\mathbf{k}+\Delta k_q\hat{\mathbf{q}})}}{\bra{n''\mathbf{k}}P\ket{n'(\mathbf{k}+\Delta k_q\hat{\mathbf{q}})}}\right.\\ & \hspace{1.5in}\left. +\ln \frac{\bra{n''\mathbf{k}}P\ket{n'\mathbf{k}}}{\iprod{n''\mathbf{k}}{n''(\mathbf{k}+\Delta k_q\hat{\mathbf{q}})}\bra{n''(\mathbf{k}+\Delta k_q\hat{\mathbf{q}})}P\ket{n'\mathbf{k}}}\right)
\end{flalign*}
each term of which is manifestly gauge invariant.  The expression can be generalized to the degenerate and near degenerate case by considering band sets and the appropriate operator sub-blocks.

\subsection{Glass Coefficient}
The solution to Maxwell's equations for and electromagnetic wave in a lossy medium can be written as $E_0\expn{i\mathbf{k}\cdot\mathbf{r}}\expn{i\omega t}$, where $\mathbf{k}$ is a complex wavevector, such  that
\begin{flalign*}
 \left(\mathbf{k}\cdot \mathbf{k}\right)= \mu \omega^2 \left( i\epsilon_i + \epsilon_r \right)
\end{flalign*}
Solving for $\mathbf{k}$(assuming propogation direction Z)
\begin{flalign*}
 k_z^2=& \mu \omega^2 \left( i\epsilon_i + \epsilon_r \right) \\
 k_z=&\pm\sqrt{\mu}\omega\left(\sqrt{\frac{\sqrt{\epsilon_r^2+\epsilon_i^2}+\epsilon_r}{2}}+i\sqrt{\frac{\sqrt{\epsilon_r^2+\epsilon_i^2}-\epsilon_r}{2}}\right)
\end{flalign*}
Also, 
\begin{flalign*}
 I\propto |E|^2=E^2_0\expn{2k_iz}\\
 I=I_0\expn{2k_iz}
\end{flalign*}
which is Beer's law.  Thus the absorption coefficient 
\begin{flalign*}
 \alpha=2k_i=\sqrt{\mu}\omega\sqrt{\frac{\sqrt{\epsilon_r^2+\epsilon_i^2}-\epsilon_r}{2}}
\end{flalign*}
can be determined from the permittivity, which from linear response theory is
\begin{flalign*}
\epsilon(\omega)=\epsilon_r(\omega)+i\epsilon_i(\omega)=1+&\frac{e}{\epsilon_0\omega^2\hbar}\sum_{mn}(f[n]-f[m])\left|\bra{m}\hat{P}\ket{n}\right|^2\mathcal{P} \frac{1}{\omega -\omega_m+\omega_n}\\ 
    &+ i\frac{e\pi}{\epsilon_0\omega^2\hbar}\sum_{mn}(f[n]-f[m])\left|\bra{m}\hat{P}\ket{n}\right|^2 \delta \left(\omega-\omega_m+\omega_n \right)
\end{flalign*}

In a lossy medium, the intensity $I$ depends on penetration depth $z$, and the total current is
\begin{flalign*}
 J_q&=w\sigma_q^{rs} \int_0^l I^{rs} \dx{z}=l\sigma\int_0^l I^{rs}_0\expn{-\alpha z}\dx{z}\\
 J_q&= w\sigma_q^{rs} I^{rs}_0\int_0^l \expn{-\alpha z}\dx{z}\\
 J_q&=w\frac{\sigma_q^{rs} I^{rs}_0}{\alpha}\left[1-\expn{-\alpha l}\right]
\end{flalign*}
where $w$ is the width of the exposed surface, and $l$ is the crystal depth.  In the limit of $\alpha l >> 1$ then
\begin{flalign}
  J_q&=w\frac{\sigma_q^{rs} I^{rs}_0}{\alpha}=wG_q^{rs}I^{rs}_0
\end{flalign}
where $G$ is the Glass coefficient~\cite{Glass74p233}.